\documentclass[prd,twocolumn,floatfix,amsmath,nofootinbib,amssymb,floatfix]{revtex4}
\usepackage{graphicx,color,dcolumn,booktabs,bm}
\usepackage{longtable,lscape}
\usepackage{pdfpages}
\usepackage{txfonts}
\usepackage{overpic}
\usepackage{amssymb}
\usepackage{makecell}
\usepackage{indentfirst}
\usepackage{feynmf}   
\usepackage{slashed}  
\usepackage{cases}
\usepackage{color}
\usepackage{multirow}
\usepackage{threeparttable}
\usepackage{epstopdf}
\usepackage{enumerate}
\usepackage{diagbox}
\usepackage{graphicx,color,dcolumn,booktabs,bm}
\usepackage[colorlinks,
            citecolor=blue,
            anchorcolor=red,
            menucolor=red,
            linkcolor=red,
            filecolor=red,
            runcolor=red,
            urlcolor=blue,
            frenchlinks=red]{hyperref}
\usepackage{tikz}
\begin{document}
\title{Investigating the spectroscopy behavior of undetected $1F$-wave charmed baryons}
\author{Si-Qiang Luo$^{1,2,3,5}$}\email{luosq15@lzu.edu.cn}
\author{Xiang Liu$^{1,3,4,5}$}\email{xiangliu@lzu.edu.cn}
\affiliation{
$^1$School of Physical Science and Technology, Lanzhou University, Lanzhou 730000, China\\
$^2$School of mathematics and statistics, Lanzhou University, Lanzhou 730000, China\\
$^3$Lanzhou Center for Theoretical Physics, Key Laboratory of Quantum Theory and Applications of the Ministry of Education, and Key Laboratory of Theoretical Physics of Gansu Province, Lanzhou University, Lanzhou 730000, China\\
$^4$MoE Frontiers Science Center for Rare Isotopes, Lanzhou University, Lanzhou 730000, China\\
$^5$Research Center for Hadron and CSR Physics, Lanzhou University and Institute of Modern Physics of CAS, Lanzhou 730000, China}

\begin{abstract}
In this work, we investigate the spectroscopic properties of $1F$-wave charmed baryons, which have not yet been observed in experiments. We employ a non-relativistic potential model and utilize the Gaussian expansion method to obtain the mass spectra of these charmed baryons. Additionally, we focus on the two-body Okubo-Zweig-Iizuka allowed strong decay behaviors, which plays a crucial role in characterizing the properties of these baryons. Our analyses of the mass spectra and two-body Okubo-Zweig-Iizuka allowed decay behaviors provides valuable insights for future experimental investigations. This study contributes to our understandings of the spectroscopic properties of $1F$-wave charmed baryons.
\end{abstract}
\maketitle

\section{Introduction}\label{sec:Introduction}

In the last two decades, there has been a growing number of observed charmed baryon states in experimental studies. These states now form the primary constituent of the current charmed baryon family listed in the Particle Data Group (PDG)~\cite{ParticleDataGroup:2020ssz}. As an integral part of the broader hadron family, the observed charmed baryon states offer an excellent platform to investigate the formation of baryons from their quark components, which is intimately linked to low-energy strong interaction phenomena (see review articles~\cite{Cheng:2015iom,Chen:2016spr,Cheng:2021qpd,Chen:2022asf}). A thorough investigation of charmed baryons can significantly enhance our understanding of the non-perturbative behavior of strong interaction.

\begin{figure}[htbp]
\centering
\includegraphics[width=8.6cm]{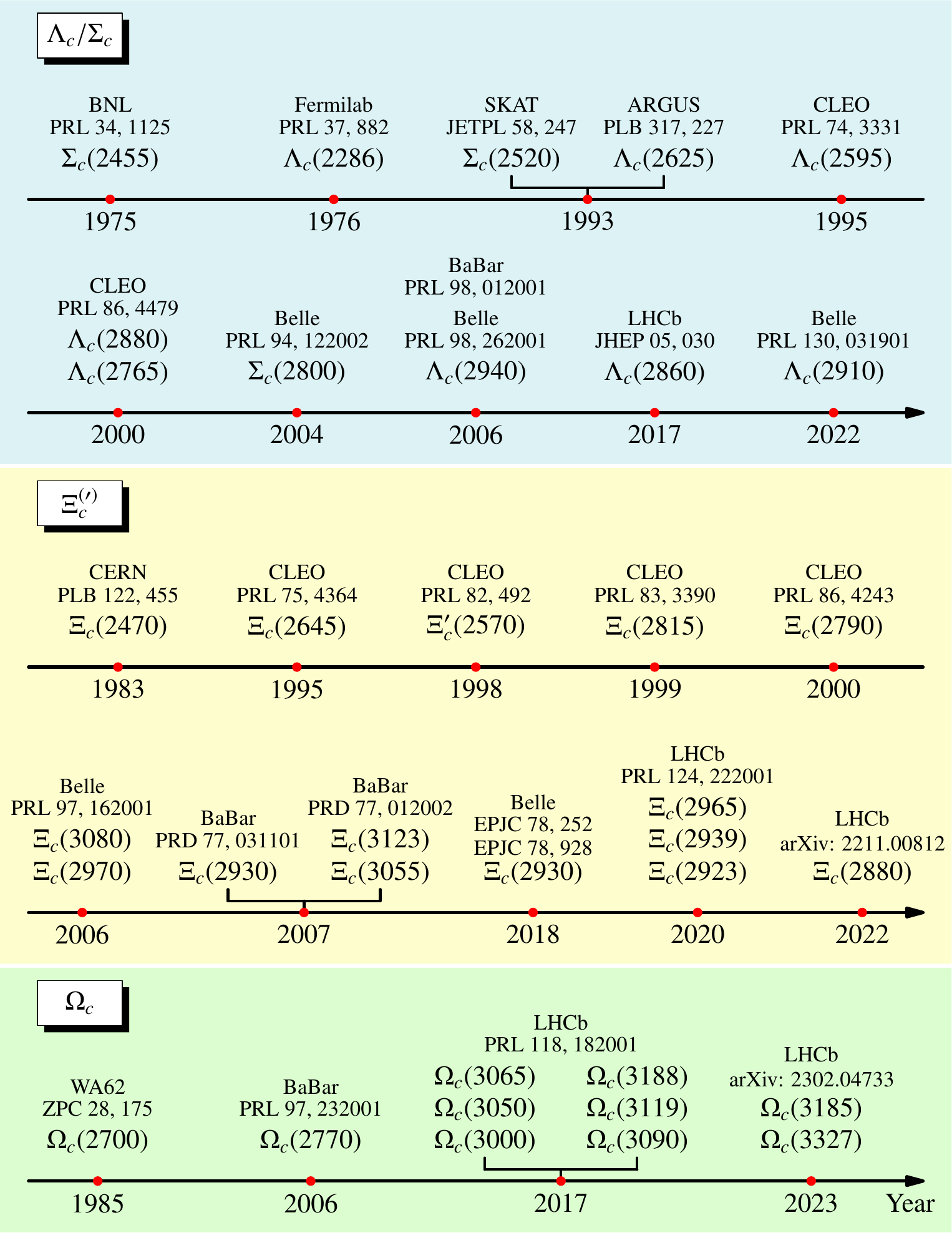}
\caption{The observed singly charmed baryons. The data presented in this paper is sourced from the following  Refs.~\cite{Cazzoli:1975et,Knapp:1976qw,Ammosov:1993pi,ARGUS:1993vtm,CLEO:1994oxm,CLEO:2000mbh,Belle:2004zjl,BaBar:2006itc,LHCb:2017jym,Belle:2022hnm,Biagi:1983en,CLEO:1998wvk,CLEO:1999msf,CLEO:2000ibb,Belle:2006edu,BaBar:2007xtc,BaBar:2007zjt,LHCb:2020iby,LHCb:2022vns,Biagi:1984mu,BaBar:2006pve,LHCb:2017uwr,LHCb:2023rtu,Belle:2017jrt,Belle:2018yob,CLEO:1995amh,Belle:2006xni}.}
\label{fig:observation}
\end{figure}

The charmed baryon, composed of a charmed quark and two light quarks, is a representative few-body system exhibiting heavy quark symmetry, which makes it relatively simple compared to light baryons. In the past near fifty years, over 30 singly charmed baryons were observed in experiments~\cite{Cazzoli:1975et,Knapp:1976qw,Ammosov:1993pi,ARGUS:1993vtm,CLEO:1994oxm,CLEO:2000mbh,Belle:2004zjl,BaBar:2006itc,LHCb:2017jym,Belle:2022hnm,Biagi:1983en,CLEO:1998wvk,CLEO:1999msf,CLEO:2000ibb,Belle:2006edu,BaBar:2007xtc,BaBar:2007zjt,LHCb:2020iby,LHCb:2022vns,Biagi:1984mu,BaBar:2006pve,LHCb:2017uwr,LHCb:2023rtu,Belle:2017jrt,Belle:2018yob,CLEO:1995amh,CLEO:1996zcj,Belle:2006xni,Solovieva:2008fw,BaBar:2008get,Belle:2017ext}. Figure \ref{fig:observation} presents the observed charmed baryon states. Intriguingly, treating the charmed baryon as a quasi two-body system by clustering the two light quarks leads to a mass spectrum that aligns with current experimental observations. The measured masses and widths allows readers to easily identify that the majority of $S$-, $P$-, and $D$-wave states are well established, indicating significant progress in hadron spectroscopy exploration~\cite{Albertus:2005zy,Chen:2007xf,Romanets:2012hm,Chen:2016iyi,Chen:2017aqm,Chen:2017gnu,Chen:2018orb,Chen:2018vuc,Chen:2019ywy,Lu:2018utx,Liang:2019aag,Lu:2019rtg,Lu:2020ivo,Liang:2020kvn,Xiao:2017dly,Zhao:2016qmh,Zhao:2017fov,Ye:2017yvl,Ye:2017dra,Zhao:2020tpf,Guo:2019ytq,Yang:2018lzg,Wang:2020gkn,Ortiz-Pacheco:2023kjn,Bijker:2020tns,Santopinto:2018ljf,Nieves:2019nol}. This achievement is a result of the join efforts of experimental and theoretical colleagues. We express our gratitude to all involved. In fact, the construction of the charms baryon family is an ongoing story. Very recently, the LHC Collaboration made an exciting announcement regarding the observation of two enhancement structures, $\Omega_c(3185)$ and $\Omega_c(3327)$, in the $\Xi_c^+ K^-$ invariant mass spectrum~\cite{LHCb:2023rtu}. The discovery of $\Omega_c(3327)$ is particularly significant as it promotes to explore the construction of $D$-wave charmed baryons~\cite{Yu:2023bxn,Luo:2023sra,Wang:2023wii}. As the LHC's high-luminosity updates continue, it is reasonable to expect that experimental access to more $D$-wave and higher orbital excitations of charmed baryons will become available. In light of these developments, theorists should devote greater attention to higher states of charmed baryon.

In this study, we investigate the spectroscopic properties of $1F$-wave charmed baryons, which have yet to be observed in experiments. We employ a non-relativistic potential model~\cite{Luo:2019qkm,Luo:2021dvj,Luo:2023sra} and utilize the Gaussian expansion method (GEM)~~\cite{Hiyama:2003cu} to obtain their mass spectrum. While the mass spectrum of $F$-wave charmed baryons is informative for future experimental searches, their two-body Okubo-Zweig-Iizuka (OZI) allowed strong decay behavior is even more crucial for characterizing their properties. To address this, we utilize the quark pair creation (QPC)~\cite{Micu:1968mk,LeYaouanc:1972vsx,LeYaouanc:1973ldf,Barnes:2007xu,Ackleh:1996yt} model, a well-established approach for analyzing the two-body OZI-allowed strong decay of singly charmed baryons~\cite{Chen:2007xf,Chen:2016iyi,Chen:2017aqm,Chen:2017gnu,Chen:2018orb,Chen:2018vuc,Chen:2019ywy,Lu:2018utx,Liang:2019aag,Lu:2019rtg,Lu:2020ivo,Liang:2020kvn,Xiao:2017dly,Zhao:2016qmh,Zhao:2017fov,Ye:2017yvl,Ye:2017dra,Zhao:2020tpf,Guo:2019ytq,Yang:2018lzg}. By employing this approach, we can estimate their total decay width, which is approximately equal to their widths. Through our analysis of the mass spectrum and two-body OZI-allowed decay behavior of $1F$-wave charmed baryons, we can provide valuable insights for future experimental investigations.

This paper is structured as follows. Following the Introduction, we proceed to present the mass spectrum of $1F$ excited singly charmed baryons in Section~\ref{sec:MassSpectrum}. Subsequently, in Section~\ref{sec:Decays}, we perform calculations for the total and partial OZI-allowed two-body strong decay widths. Finally, we conclude with a concise summary in Section~\ref{sec:Summary}.

\section{Mass spectrum}\label{sec:MassSpectrum}

As the first step, we utilize a non-relativistic potential model to accurately compute the mass spectra of the $F$-wave excited singly charmed baryons. The Hamiltonian~\cite{Luo:2019qkm,Luo:2021dvj,Luo:2023sra} is
\begin{equation}\label{eq:H}
\hat{H}=\sum\limits_{i}\left(m_i+\frac{p_i^2}{2m_i}\right)+\sum\limits_{i<j}\left(V_{ij}^{\rm conf}+V_{ij}^{\rm hyp}+V_{ij}^{\rm so(cm)}+V_{ij}^{\rm so(tp)}\right).
\end{equation}
Here, we denote the mass and momentum of the $i$th constituent quark as $m_i$ and $p_i$, respectively. The terms $V_{ij}^{\rm conf}$, $V_{ij}^{\rm hyp}$, $V_{ij}^{\rm so(cm)}$, and $V_{ij}^{\rm so(tp)}$ in Eq.~(\ref{eq:H}) represent the confinement, hyperfine, color-magnetic, and Thomas-precession potentials, respectively. The specific forms of these interactions can be expressed as follows:
\begin{equation}\label{eq:Vconf}
\begin{split}
V_{ij}^{\rm conf}=-\frac{2}{3}\frac{\alpha_s}{r_{ij}}+\frac{b}{2}r_{ij}+\frac{1}{2}C,
\end{split}
\end{equation}
\begin{equation}\label{eq:Vhyp}
\begin{split}
V_{ij}^{\rm hyp}=&\frac{2\alpha_s}{3m_im_j}\left[\frac{8\pi}{3}\tilde{\delta}(r_{ij}){\bf s}_i\cdot{\bf s}_j+\frac{1}{r_{ij}^3}S({\bf r},{\bf s}_i,{\bf s}_j)\right],\\
\tilde{\delta}(r)=&\frac{\sigma^3}{\pi^{3/2}}{\rm e}^{-\sigma^2r^2},~~~
S({\bf r},{\bf s}_i,{\bf s}_j)=\frac{3{\bf s}_i\cdot{\bf r}_{ij}{\bf s}_j\cdot{\bf r}_{ij}}{r_{ij}^2}-{\bf s}_i\cdot{\bf s}_j,
\end{split}
\end{equation}
\begin{equation}\label{eq:Vsocm}
\begin{split}
V_{ij}^{{\rm so(cm)}}=&\frac{2\alpha_s}{3r_{ij}^3}\left(\frac{{\bf r}_{ij}\times{\bf p}_i\cdot{\bf s}_i}{m_i^2}-\frac{{\bf r}_{ij}\times{\bf p}_j\cdot{\bf s}_j}{m_j^2}\right.\\
&\left.-\frac{{\bf r}_{ij}\times{\bf p}_j\cdot{\bf s}_i-{\bf r}_{ij}\times{\bf p}_i\cdot{\bf s}_j}{m_im_j}\right),
\end{split}
\end{equation}
\begin{equation}\label{eq:Vsotp}
V_{ij}^{{\rm so(tp)}}=-\frac{1}{2r_{ij}}\frac{\partial H_{ij}^{\rm conf}}{\partial r_{ij}}\left(\frac{{\bf r}_{ij}\times{\bf p}_i\cdot{\bf s}_i}{m_i^2}-\frac{{\bf r}_{ij}\times{\bf p}_j\cdot{\bf s}_j}{m_j^2}\right).
\end{equation}
In Eqs.~(\ref{eq:Vconf})-(\ref{eq:Vsotp}), the parameters $\alpha_s$, $b$, $C$, and $\sigma$ represent the coupling constant of the one-gluon exchange, the strength of the linear confinement, the renormalized mass constant, and the smearing parameter, respectively.

\begin{table}[htbp]
\caption{The parameters involved in the adopted potential model.}
\label{tab:parameter}
\renewcommand\arraystretch{1.25}
\begin{tabular*}{86mm}{@{\extracolsep{\fill}}m{10mm}m{15mm}<{\centering}m{15mm}<{\centering}m{15mm}<{\centering}m{15mm}<{\centering}}
\toprule[1.00pt]
\toprule[1.00pt]
System               &$\alpha_s$ &$b$ (GeV$^2$) &$\sigma$ (GeV) &$C$ (GeV)  \\
\midrule[0.75pt]
$\Lambda_c/\Sigma_c$ &0.560      &0.122         &1.600          &$-$0.633   \\
$\Xi_c^{(\prime)}$   &0.560      &0.140         &1.600          &$-$0.693   \\
$\Omega_c$           &0.578      &0.144         &1.732          &$-$0.688   \\
meson                &0.578      &0.144         &1.028          &$-$0.685   \\
\midrule[0.75pt]
\multicolumn{5}{c}{\mbox{$m_{u/d}=0.370~{\rm GeV}~m_{s}=0.600~{\rm GeV}~m_{c}=1.880~{\rm GeV}$}}\\
\bottomrule[1.00pt]
\bottomrule[1.00pt]
\end{tabular*}
\end{table}

\begin{table}[htbp]
\centering
\caption{The basis of $\lambda$-mode excited $1F$ singly charmed baryons.}
\label{tab:basis}
\renewcommand\arraystretch{1.25}
\begin{tabular*}{86mm}{@{\extracolsep{\fill}}clcccccccc}
\toprule[1.00pt]
\toprule[1.00pt]
Symmetry                                             
&States                                              &$J$           &$s_\ell$ &$n_\rho$ &$n_\lambda$ &$l_\rho$ &$l_\lambda$ &$L$ &$j_\ell$ \\
\midrule[0.75pt]
\multirow{2}{*}{$\bar{3}_f$}
&$\Lambda_c/\Xi_c(1F,5/2^-)$                         &$\frac{5}{2}$ &0        &0        &0           &0        &3           &3   &3        \\
&$\Lambda_c/\Xi_c(1F,7/2^-)$                         &$\frac{7}{2}$ &0        &0        &0           &0        &3           &3   &3        \\
\midrule[0.75pt]
\multirow{6}{*}{$6_f$}
&$\Sigma_{c2}/\Xi^\prime_{c2}/\Omega_{c2}(1F,3/2^-)$ &$\frac{3}{2}$ &1        &0        &0           &0        &3           &3   &2        \\
&$\Sigma_{c2}/\Xi^\prime_{c2}/\Omega_{c2}(1F,5/2^-)$ &$\frac{5}{2}$ &1        &0        &0           &0        &3           &3   &2        \\
&$\Sigma_{c3}/\Xi^\prime_{c3}/\Omega_{c3}(1F,5/2^-)$ &$\frac{5}{2}$ &1        &0        &0           &0        &3           &3   &3        \\
&$\Sigma_{c3}/\Xi^\prime_{c3}/\Omega_{c3}(1F,7/2^-)$ &$\frac{7}{2}$ &1        &0        &0           &0        &3           &3   &3        \\
&$\Sigma_{c4}/\Xi^\prime_{c4}/\Omega_{c4}(1F,7/2^-)$ &$\frac{7}{2}$ &1        &0        &0           &0        &3           &3   &4        \\
&$\Sigma_{c4}/\Xi^\prime_{c4}/\Omega_{c4}(1F,9/2^-)$ &$\frac{9}{2}$ &1        &0        &0           &0        &3           &3   &4        \\
\bottomrule[1.00pt]
\bottomrule[1.00pt]
\end{tabular*}
\end{table}

\begin{table*}
\centering
\caption{A comparison of predicted masses for $F$-wave singly charmed baryons from various studies. Here, the listed masses are in units of MeV.}
\label{tab:massspectrum}
\renewcommand\arraystretch{1.15}
\begin{tabular*}{178mm}{@{\extracolsep{\fill}}lccclccclccc}
\toprule[1.00pt]
\toprule[1.00pt]
States                  &Our  &Ref.~\cite{Ebert:2011kk} &Ref.~\cite{Yu:2022ymb} &States                      &Our  &Ref.~\cite{Ebert:2011kk} &Ref.~\cite{Li:2022xtj} &States                  &Our  &Ref.~\cite{Ebert:2011kk} &Ref.~\cite{Yu:2022ymb} \\
\midrule[0.75pt]
$\Lambda_c(1F,5/2^-)$   &3075 &3097                     &3104                   &$\Xi_c(1F,5/2^-)$           &3292 &3278                     &3289                   &                        &     &                         &                       \\
$\Lambda_c(1F,7/2^-)$   &3079 &3078                     &3111                   &$\Xi_c(1F,7/2^-)$           &3295 &3292                     &3294                   &                        &     &                         &                       \\
$\Sigma_{c2}(1F,3/2^-)$ &3276 &3288                     &3299                   &$\Xi_{c2}^\prime(1F,3/2^-)$ &3427 &3418                     &3424                   &$\Omega_{c2}(1F,3/2^-)$ &3540 &3533                     &3525                   \\
$\Sigma_{c2}(1F,5/2^-)$ &3283 &3254                     &3304                   &$\Xi_{c2}^\prime(1F,3/2^-)$ &3433 &3394                     &3428                   &$\Omega_{c2}(1F,5/2^-)$ &3547 &3515                     &3528                   \\
$\Sigma_{c3}(1F,5/2^-)$ &3247 &3283                     &3299                   &$\Xi_{c3}^\prime(1F,3/2^-)$ &3408 &3408                     &3424                   &$\Omega_{c3}(1F,5/2^-)$ &3532 &3522                     &3525                   \\
$\Sigma_{c3}(1F,7/2^-)$ &3252 &3227                     &3305                   &$\Xi_{c3}^\prime(1F,3/2^-)$ &3412 &3373                     &3428                   &$\Omega_{c3}(1F,7/2^-)$ &3537 &3498                     &3529                   \\
$\Sigma_{c4}(1F,7/2^-)$ &3207 &3253                     &3299                   &$\Xi_{c4}^\prime(1F,3/2^-)$ &3382 &3393                     &3423                   &$\Omega_{c4}(1F,7/2^-)$ &3521 &3514                     &3524                   \\
$\Sigma_{c4}(1F,9/2^-)$ &3209 &3209                     &3305                   &$\Xi_{c4}^\prime(1F,3/2^-)$ &3383 &3357                     &3428                   &$\Omega_{c4}(1F,9/2^-)$ &3520 &3485                     &3529                   \\
\bottomrule[1.00pt]
\bottomrule[1.00pt]
\end{tabular*}
\end{table*}

As a conventional three-body system, it is advantageous to employ the $\rho$- and $\lambda$-modes to represent the Jacobi coordinates of a singly heavy baryon. In this context, the $\rho$-mode corresponds to the coordinate between the two light flavor quarks $q_1$ and $q_2$ ($q=u,\,d,\,s$). On the other hand, the $\lambda$-mode represents the vector connecting the heavy flavor quark $Q_3$ ($Q$=$c,\,b$) to the center-of-mass of the two light flavor quarks. The basis employed to derive the mass spectrum of the aforementioned $1F$-wave charmed baryons represents
\begin{equation}\label{eq:basis}
|JM\rangle=|[[s_{q_1}s_{q_2}]_{s_\ell}[n_\rho n_\lambda l_\rho l_\lambda]_L]_{j_\ell}s_{Q_3}]_{JM}\rangle.
\end{equation}
In this context, $s_{q_1}$, $s_{q_2}$, and $s_{Q_3}$ represent the spins of the quarks involved. Meanwhile, $s_{\ell}$ and $j_{\ell}$ denote the spin and total angular momentum of the light degree of freedom, respectively. The quantum numbers $n_{\rho/\lambda}$ and $l_{\rho/\lambda}$ refer to the radial and orbital components, respectively. The total orbital angular momentum of the system is denoted as $L$. Based on the spectroscopy of observed charmed baryons, these states can be categorized as $\lambda$-mode excitations. In this study, our primary focus remains on the $\lambda$-mode excited $1F$ states. Besides the degree-of-freedom in spin-spatial, a singly charmed baryon also has flavor wave function. Within the framework of $SU(3)$ flavor symmetry, the coupling of flavor wave functions can be decomposed as $3\otimes 3=\bar{3}\oplus 6$. The states in $\bar{3}_f$ include $\Lambda_c^+$, $\Xi_c^+$, and $\Xi_c^0$, while the $6_f$ states consist of $\Sigma_c^+$, $\Sigma_c^0$, $\Sigma_c^{++}$, $\Xi_c^{\prime+}$, $\Xi_c^{\prime 0}$, and $\Omega_c^0$. It is worth noting that the flavor wave functions of the two light quarks are anti-symmetrical for $\bar{3}_f$ and symmetrical for $6_f$. For the sake of simplicity, we will omit their isospin partners in the subsequent discussions. Adhering to the Pauli principle, we present the basis states in Table~\ref{tab:basis}. The $\lambda$-mode excited $1F$ states of $\Lambda_c$ or $\Xi_c$ consist of two states, while the $\lambda$-mode excited $1F$ states of $\Sigma_c$, $\Xi_c^\prime$, or $\Omega_c$ encompass six states.

\begin{figure*}
\begin{center}
\includegraphics[width=17.8cm,keepaspectratio]{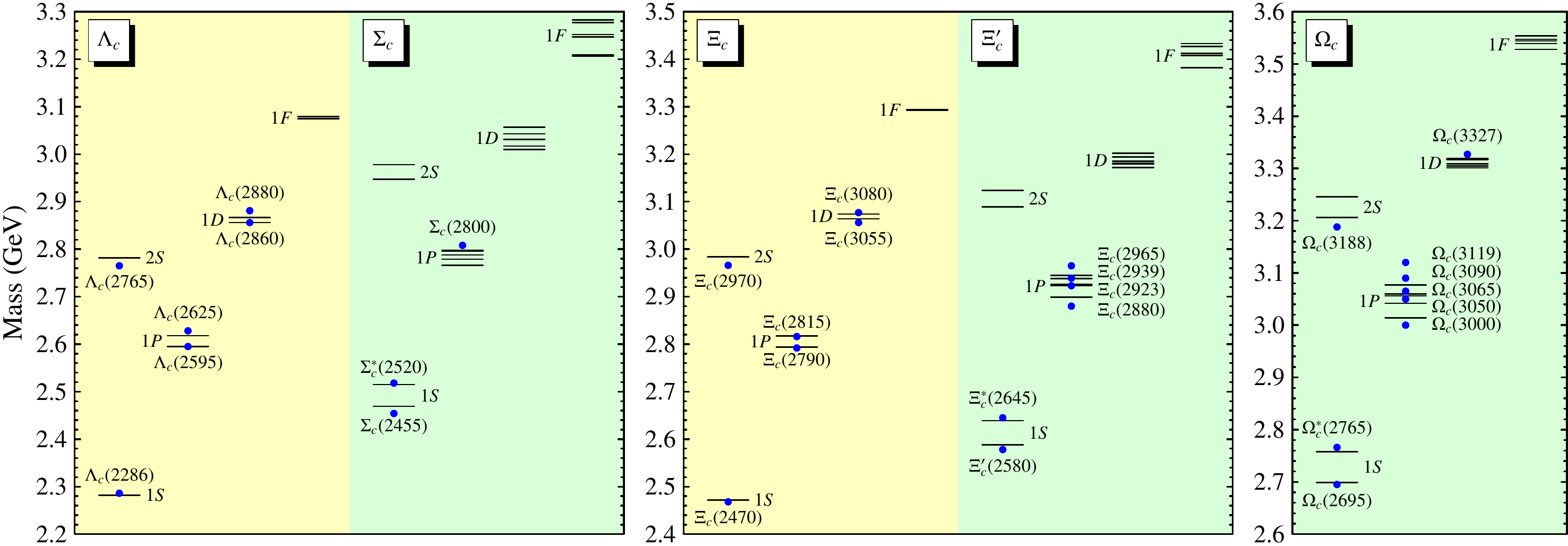}
\caption{The calculated masses of the singly charmed baryons and the comparison with experimental data. The short lines in the graph represent the calculated results, while the blue points on the graph are obtained from the experimental data, which is taken from the PDG~\cite{ParticleDataGroup:2020ssz}.}
\label{fig:spectrum}
\end{center}
\end{figure*}

Using the Gaussian expansion method~\cite{Hiyama:2003cu}, we can solve the three body Schr\"odinger equations with the above potentials. According to the mass spectra of the observed singly charmed baryons, we have derived the parameters of the potential model and compiled them in Table \ref{tab:parameter}. Additionally, we have compared our calculated results with the masses of the observed states, as depicted in Figure \ref{fig:spectrum}. Our calculations for the $1S$, $2S$, $1P$, and $1D$ states exhibit a remarkable agreement with the corresponding experimental candidates. Similar results were also obtained in previous theoretical works~\cite{Ebert:2011kk,Yu:2022ymb,Yoshida:2015tia,Roberts:2007ni,Yamaguchi:2014era,Shah:2016nxi,Garcia-Tecocoatzi:2022zrf,Mao:2017wbz,Li:2022xtj}. Building upon these initial considerations, we have proceeded to calculate the masses of the $\lambda$-mode $1F$ excited singly charmed baryons. In accordance with the Pauli principle, we present the basis states in Table \ref{tab:basis}. The $\lambda$-mode excited $1F$ states of $\Lambda_c$ or $\Xi_c$ consist of two states, while the $\lambda$-mode excited $1F$ states of $\Sigma_c$, $\Xi_c^\prime$, or $\Omega_c$ encompass six states. The numerical results for these states are provided in Table \ref{tab:massspectrum}.

\section{OZI-allowed two-body strong decays}\label{sec:Decays}

Since the decay behavior can provide valuable information in the search for these states, it is essential to perform systematic calculations for the decay widths. In this study, we utilize the QPC model~\cite{Micu:1968mk,LeYaouanc:1972vsx,LeYaouanc:1973ldf,Barnes:2007xu,Ackleh:1996yt} to calculate the partial and total decay widths of $\lambda$-mode $1F$ excited singly charmed baryons. The corresponding transition operator is used in these calculations, i.e.,
\begin{equation}\label{eq:toperator}
\begin{split}
\hat{\cal T}=&-3\gamma\sum_{m}\langle 1,m;1,-m|0,0\rangle\int{\rm d}^3{\bf p}_i\;{\rm d}^3{\bf p}_j\;\delta({\bf p}_i+{\bf p}_j)\\
          &\times\mathcal{Y}_1^m\left(\frac{{\bf p}_i-{\bf p}_j}{2}\right)\omega_0^{(i,j)}\phi_0^{(i,j)}\chi_{1,-m}^{(i,j)}b^\dagger_i({\bf p}_i)d^\dagger_j({\bf p}_j),
\end{split}
\end{equation}
where $\omega_0^{(i,j)}=(r\bar{r}+g\bar{g}+b\bar{b})/\sqrt{3}$, $\phi_0^{(i,j)}=(u\bar{u}+d\bar{d}+s\bar{s})/\sqrt{3}$, $\chi_{1,-m}^{(i,j)}$, and $\mathcal{Y}$ are the color-singlet, flavor-singlet, spin-1, and spatial functions for the quark pair created from the vacuum, respectively. The $\gamma$ is a dimensionless parameter, which is determined by the widths of the well established states. Then in the decay process $A\to BC$, the partial wave amplitude could be written as
\begin{equation}\label{eq:MAtoBC}
{\cal M}_{A\to BC}^{L_{BC}S_{BC}}(p)=\langle BC,L_{BC}S_{BC},p|\hat{\cal T}|A\rangle,
\end{equation}
where $L_{BC}$ and $S_{BC}$ are the relative orbital angular momentum and spin between the final $BC$. In Eq.~(\ref{eq:MAtoBC}), the variable $p$ represents the momentum of the outgoing baryon $B$. The partial decay width can be obtained by performing the following calculation:
\begin{equation}
\Gamma_{A\to BC}=2\pi\frac{E_BE_C}{M_A}p\sum\limits_{L_{BC}S_{BC}}|{\cal M}_{A\to BC}^{L_{BC}S_{BC}}(p)|^2,
\end{equation}
where $E_B=\sqrt{M_B^2+p^2}$ and $E_C=\sqrt{M_C^2+p^2}$ are the energies of the baryon $B$ and meson $C$, respectively.

To calculate the overlap of the spatial part in Eq.~(\ref{eq:MAtoBC}), we utilize the following reduced wave functions:
\begin{equation}\label{eq:spatialfun}
\begin{split}
\psi_{n_\rho n_\lambda l_\rho l_\lambda LM}({\bm \rho},{\bm \lambda})\approx&R^{p}_{n_\rho l_\rho}(\beta_\rho,p_\rho) R^{p}_{n_\lambda l_\lambda}(\beta_\lambda,p_\lambda)\\
&\times  \sum\limits_{m_\rho m_\lambda}C_{l_\rho m_\rho,l_\lambda m_\lambda}^{LM}Y_{l_\rho m_\rho}(\Omega_{{\bf p}_\rho}) Y_{l_\lambda m_\lambda}(\Omega_{{\bf p}_\lambda}),
\end{split}
\end{equation}
where
\begin{equation}\label{eq:sho}
\begin{split}
R^p_{nl}(\beta,P)=&\frac{(-1)^n(-{\mathrm i})^l}{\beta^{\frac{3}{2}+l}}\sqrt{\frac{2n!}{\Gamma(n+l+\frac{3}{2})}}L_{n}^{l+\frac{1}{2}}({P^2}/{\beta^2}){\mathrm e}^{-\frac{P^2}{2\beta^2}}P^l
\end{split}
\end{equation}
is radial part of the simple harmonic oscillator (SHO) wave function. In Eqs.~(\ref{eq:spatialfun})-(\ref{eq:sho}), the $\beta$ is a characteristic parameter to represent the simple harmonic oscillator. Using the approaches described in Refs. \cite{Chen:2016iyi,Chen:2018orb,Luo:2023sra}, we determine the values of $\beta$ and present them in Table~\ref{tab:beta}.

Based on the obtained parameters, we determine the value of the parameter $\gamma=9.58$ from the measured width of $\Sigma_c^*(2520)$~\cite{ParticleDataGroup:2020ssz}. This value of $\gamma$ is employed in the global calculations conducted in this study.

\begin{table}
\centering
\caption{The $\beta$ values used in this work.}
\label{tab:beta}
\renewcommand\arraystretch{1.25}
\begin{tabular*}{86mm}{@{\extracolsep{\fill}}lcclcclc}
\toprule[1.00pt]
\toprule[1.00pt]
States           &$\beta_\rho$ &$\beta_\lambda$ &States             &$\beta_\rho$ &$\beta_\lambda$ &States &$\beta$ \\
\midrule[0.75pt]
$\Lambda_c(1S)$  &0.290        &0.344           &$\Xi_c(1S)$        &0.301        &0.383           &$\pi$  &0.409   \\
$\Lambda_c(2S)$  &0.251        &0.185           &$\Xi_c(2S)$        &0.258        &0.207           &$K$    &0.385   \\
$\Lambda_c(1P)$  &0.271        &0.237           &$\Xi_c(1P)$        &0.281        &0.265           &$K^*$  &0.259   \\
$\Lambda_c(2P)$  &0.256        &0.153           &$\Xi_c(2P)$        &0.264        &0.173           &$D$    &0.357   \\
$\Lambda_c(1D)$  &0.259        &0.182           &$\Xi_c(1D)$        &0.268        &0.203           &$D^*$  &0.307   \\
$\Lambda_c(1F)$  &0.254        &0.152           &$\Xi_c(1F)$        &0.262        &0.167           &       &        \\
$\Sigma_c(1S)$   &0.220        &0.336           &$\Xi_c^\prime(1S)$ &0.252        &0.383           &       &        \\
$\Sigma_c^*(1S)$ &0.212        &0.315           &$\Xi_c^*(1S)$      &0.243        &0.358           &       &        \\
$\Sigma_c(2S)$   &0.188        &0.186           &$\Xi_c^\prime(2S)$ &0.212        &0.210           &       &        \\
$\Sigma_c^*(2S)$ &0.190        &0.174           &$\Xi_c^*(2S)$      &0.216        &0.202           &       &        \\
$\Sigma_c(1P)$   &0.210        &0.238           &$\Xi_c^\prime(1P)$ &0.240        &0.270           &       &        \\
$\Sigma_c(1D)$   &0.198        &0.185           &$\Xi_c^\prime(1D)$ &0.226        &0.206           &       &        \\
$\Sigma_c(1F)$   &0.191        &0.152           &$\Xi_c^\prime(1F)$ &0.218        &0.168           &       &        \\
$\Omega_c(1S)$   &0.288        &0.420           &$N$                &0.280        &0.324           &       &        \\
$\Omega_c^*(1S)$ &0.275        &0.389           &$\Delta$           &0.249        &0.288           &       &        \\
$\Omega_c(2S)$   &0.230        &0.229           &$\Lambda$          &0.281        &0.285           &       &        \\
$\Omega_c^*(2S)$ &0.236        &0.217           &$\Sigma$           &0.223        &0.301           &       &        \\
$\Omega_c(1P)$   &0.273        &0.294           &$\Sigma^*$         &0.206        &0.262           &       &        \\
$\Omega_c(1D)$   &0.254        &0.223           &$\Xi$              &0.287        &0.317           &       &        \\
$\Omega_c(1F)$   &0.244        &0.181           &$\Xi^*$            &0.258        &0.265           &       &        \\
\bottomrule[1.00pt]
\bottomrule[1.00pt]
\end{tabular*}
\end{table}

\subsection{$\Lambda_c(1F)$ states}

\begin{table}
\centering
\caption{The partial and total widths of the $\Lambda_c(1F)$ in units of MeV. The $M_f$ represents the masses of the final singly charmed baryons. In the table, channels with small partial widths are listed in the column labeled "..." to indicate their negligible contribution to the overall decay process.}
\label{tab:Lambdac1F}
\renewcommand\arraystretch{1.15}
\begin{tabular*}{86mm}{@{\extracolsep{\fill}}lccc}
\toprule[1.00pt]
\toprule[1.00pt]
Decay channels             &$M_f$ (MeV) &$\Lambda_c(1F,5/2^-)$ &$\Lambda_c(1F,7/2^-)$ \\
\midrule[0.75pt]
$\Sigma_c(1S,3/2^+)\pi$    &2520        & 0.5                  & 0.8                  \\
$\Sigma_{c2}(1P,3/2^-)\pi$ &2779        & 9.5                  & 0.2                  \\
$\Sigma_{c2}(1P,5/2^-)\pi$ &2796        & 0.8                  & 9.5                  \\
$N D$                      &            & 9.9                  &11.8                  \\
$N D^*$                    &            &21.6                  &40.2                  \\
$\cdots$                   &            & 1.0                  & 0.8                  \\
\midrule[0.75pt]
Total                      &            &43.3                  &63.3                  \\
\bottomrule[1.00pt]
\bottomrule[1.00pt]
\end{tabular*}
\end{table}

Table~\ref{tab:Lambdac1F} presents the calculated widths of $\Lambda_c(1F,5/2^-)$ and $\Lambda_c(1F,7/2^-)$ to be approximately 43.3 MeV and 63.3 MeV, respectively. It is worth noting that we observe a significant contribution from the $ND^*$ decay channel in both states, as indicated by the large branch ratios, i.e.,
\begin{equation}
\begin{split}
{\rm Br}~[\Lambda_c(1F,5/2^-)\to ND^*]\approx&49.9\%,\\
{\rm Br}~[\Lambda_c(1F,7/2^-)\to ND^*]\approx&63.5\%.\\
\end{split}
\end{equation}

The observation of the $ND^*$ channel in the calculated widths of $\Lambda_c(1F,5/2^-)$ and $\Lambda_c(1F,7/2^-)$ is noteworthy and may have connections to previous experimental observations. The LHCb Collaboration has reported the observation of two $\Lambda_c(1D)$ candidates, the $\Lambda_c(2860)$ and $\Lambda_c(2880)$, as well as a $\Lambda_c(2P)$ candidate, the $\Lambda_c(2940)$, in the $\Lambda_b^0\to\Lambda_c^+(X)\pi^-\to D^0 p\pi^-$ decay channel~\cite{LHCb:2017jym}. Given that the $\Lambda_c(1F)$ states exhibit similar excited modes to the $\Lambda_c(1D)$ and $\Lambda_c(2P)$ states, experimentalists could search for $\Lambda_c(1F)$ states in the chain process $\Lambda_b^0\to \Lambda_c^+(1F)\pi^-\to D^{*0}p\pi^-$. This approach may provide valuable insights into the existence and properties of $\Lambda_c(1F)$ states.

\subsection{$\Xi_c(1F)$ states}

\begin{table}
\centering
\caption{The partial and total widths of the $\Xi_c(1F)$ in units of MeV. In the calculation of partial decay widths, the $M_f$ denotes the masses of the final singly charmed baryons. The column marked with ``$\cdots$" includes channels with small partial widths. A value of "0.0" indicates that the width is less than 0.1 MeV.}
\label{tab:Xic1F}
\renewcommand\arraystretch{1.15}
\begin{tabular*}{86mm}{@{\extracolsep{\fill}}lccc}
\toprule[1.00pt]
\toprule[1.00pt]
Decay channels                 &$M_f$ (MeV) &$\Xi_c(1F,5/2^-)$ &$\Xi_c(1F,7/2^-)$ \\
\midrule[0.75pt]
$\Xi^\prime_{c2}(1P,3/2^-)\pi$ &2926        & 1.5              & 0.1              \\
$\Xi^\prime_{c2}(1P,5/2^-)\pi$ &2945        & 0.2              & 1.6              \\
$\Sigma_c(1S,1/2^+)\bar{K}$    &2455        & 0.7              & 0.7              \\
$\Sigma_c(1S,3/2^+)\bar{K}$    &2520        & 1.2              & 1.7              \\
$\Sigma_{c2}(1P,3/2^-)\bar{K}$ &2779        & 4.4              & 0.0              \\
$\Sigma_{c2}(1P,5/2^-)\bar{K}$ &2796        & 0.0              & 0.6              \\
$\Lambda D$                    &            & 0.5              & 2.1              \\
$\Sigma D$                     &            &10.0              &22.9              \\
$\Lambda D^*$                  &            & 4.0              & 5.2              \\
$\Sigma D^*$                   &            &28.3              &54.3              \\
$\cdots$                       &            & 0.9              & 0.9              \\
\midrule[0.75pt]
Total                          &            &51.7              &90.1              \\
\bottomrule[1.00pt]
\bottomrule[1.00pt]
\end{tabular*}
\end{table}

In the realm of orbital excited $\Xi_c$ states, the established candidates include the $1P$ and $1D$ states. However, the $\Xi_c(1F)$ states have not been observed yet. In this study, we provide predictions for the masses and widths of the $\Xi_c(1F)$ states, which could be instrumental in the search for higher orbital excited $\Xi_c$ states by experimentalists. As shown in Table~\ref{tab:Xic1F}, the total widths of $\Xi_c(1F,5/2^-)$ and $\Xi_c(1F,7/2^-)$ are projected to be 51.7 and 90.1 MeV, respectively. These results indicate that the $\Xi_c(1F,7/2^-)$ state is expected to exhibit large widths but the $\Xi_c(1F,5/2^-)$ state is not so broad. Furthermore, it is worth exploring the $\Sigma D$ and $\Sigma D^*$ channels as potential avenues for the observation of the $\Xi_c(1F)$ states. For the $\Sigma D^*$ channel, the calculated branch ratios of both $\Xi_c(1F,5/2^-)$ and $\Xi_c(1F,7/2^-)$ are larger than 50\%. We recommend conducting a search for the $\Xi_c(1F)$ states in the aforementioned channels.

\subsection{$\Sigma_c(1F)$ states}

\begin{table*}
\centering
\caption{The partial and total widths of the $\Sigma_c(1F)$ in units of MeV. The $M_f$ corresponds to the masses of the final singly charmed baryons. Channels with small partial widths are marked with "..." in the respective column to indicate their negligible contribution. A value of "0.0" indicates that the partial width is less than 0.1 MeV. The symbol "$\times$" is used to denote that the coupling is forbidden. If the mass of a initial state is below the threshold for a particular decay channel, it is denoted by "-".}
\label{tab:Sigmac1F}
\renewcommand\arraystretch{1.15}
\begin{tabular*}{178mm}{@{\extracolsep{\fill}}lccccccc}
\toprule[1.00pt]
\toprule[1.00pt]
Decay channels             &$M_f$ (MeV) &$\Sigma_{c2}(1F,3/2^-)$ &$\Sigma_{c2}(1F,5/2^-)$ &$\Sigma_{c3}(1F,5/2^-)$ &$\Sigma_{c3}(1F,7/2^-)$ &$\Sigma_{c4}(1F,7/2^-)$ &$\Sigma_{c4}(1F,9/2^-)$ \\
\midrule[0.75pt]
$\Lambda_c(2S,1/2^+)\pi$   &2766        & 7.7                    &  8.5                   &$\times$                &$\times$                & 6.5                    & 6.7                    \\
$\Lambda_c(1P,1/2^-)\pi$   &2592        & 3.5                    &  0.0                   & 1.8                    & 1.0                    & 8.1                    & 1.7                    \\
$\Lambda_c(1P,3/2^-)\pi$   &2628        & 0.5                    &  3.3                   & 2.0                    & 2.8                    & 4.4                    &10.9                    \\
$\Lambda_c(2P,1/2^-)\pi$   &3004        & 3.6                    &  0.0                   & 0.1                    & 0.1                    & 0.1                    & 0.0                    \\
$\Lambda_c(2P,3/2^-)\pi$   &2940        & 2.8                    & 15.4                   & 0.5                    & 0.8                    & 0.2                    & 0.9                    \\
$\Lambda_c(1D,3/2^+)\pi$   &2856        &14.5                    &  0.7                   &17.4                    & 3.1                    & 9.1                    & 0.4                    \\
$\Lambda_c(1D,5/2^+)\pi$   &2881        & 1.5                    & 13.1                   & 4.3                    &17.2                    & 1.2                    & 8.9                    \\
$\Lambda_c(1F,5/2^-)\pi$   &3075        &17.8                    &  1.0                   & 3.6                    & 0.2                    &$-$                     &$-$                     \\
$\Lambda_c(1F,7/2^-)\pi$   &3079        & 0.0                    & 17.8                   & 0.2                    & 3.8                    &$-$                     &$-$                     \\
$\Sigma_{c0}(1P,1/2^-)\pi$ &2788        &$\times$                &$\times$                & 3.0                    & 3.1                    &$\times$                &$\times$                \\
$\Sigma_{c1}(1P,1/2^-)\pi$ &2766        & 0.1                    &  1.0                   & 3.9                    & 2.3                    & 2.4                    & 0.1                    \\
$\Sigma_{c1}(1P,3/2^-)\pi$ &2798        & 1.2                    &  1.1                   & 4.0                    & 5.6                    & 0.7                    & 2.5                    \\
$\Sigma_{c2}(1P,3/2^-)\pi$ &2779        & 1.6                    &  1.6                   & 1.9                    & 0.5                    & 5.6                    & 1.0                    \\
$\Sigma_{c2}(1P,5/2^-)\pi$ &2796        & 2.1                    &  2.4                   & 0.6                    & 2.9                    & 1.6                    & 5.5                    \\
$\Sigma_{c2}(1D,3/2^+)\pi$ &3030        &18.3                    &  1.6                   & 1.1                    & 0.2                    & 0.2                    & 0.0                    \\
$\Sigma_{c2}(1D,5/2^+)\pi$ &3043        & 1.5                    & 19.9                   & 0.2                    & 1.0                    & 0.0                    & 0.1                    \\
$\Sigma_{c3}(1D,5/2^+)\pi$ &3010        & 3.6                    &  0.7                   &25.4                    & 0.3                    & 1.5                    & 0.1                    \\
$\Sigma_{c3}(1D,7/2^+)\pi$ &3017        & 0.6                    &  3.7                   & 0.3                    &26.2                    & 0.1                    & 1.3                    \\
$N D$                      &            & 0.0                    &  0.0                   & 0.0                    & 1.6                    & 0.1                    & 3.6                    \\
$\Delta D$                 &            & 4.8                    & 15.0                   &14.0                    & 7.4                    &23.9                    & 1.3                    \\
$N D^*$                    &            & 0.5                    &  1.0                   & 4.7                    & 3.5                    & 7.6                    & 4.3                    \\
$\Delta D^*$               &            & 1.9                    &  4.8                   & 0.1                    & 0.6                    &$-$                     &$-$                     \\
$\Sigma D_s$               &            & 2.5                    &  0.1                   & 1.0                    & 0.0                    & 0.0                    & 0.0                    \\
$\cdots$                   &            & 2.3                    &  2.2                   & 3.3                    & 3.0                    & 1.7                    & 1.7                    \\
\midrule[0.75pt]
Total                      &            &92.9                    &114.9                   &93.4                    &87.2                    &75.0                    &51.0                    \\
\bottomrule[1.00pt]
\bottomrule[1.00pt]
\end{tabular*}
\end{table*}

In Table~\ref{tab:Sigmac1F}, the calculated widths of $\Sigma_{c2}(1F,3/2^-)$, $\Sigma_{c2}(1F,5/2^-)$, $\Sigma_{c3}(1F,5/2^-)$, $\Sigma_{c3}(1F,7/2^-)$, and $\Sigma_{c4}(1F,7/2^-)$ fall within the range of approximately 70 to 110 MeV. However, the predicted width of $\Sigma_{c4}(1F,9/2^-)$ is smaller, with a value of 51.0 MeV, compared to the widths of the former five states.

Upon careful analysis of their decay modes, it is observed that the $\Lambda_c(1S)\pi$ and $\Sigma_c(1S)\pi$ channels have significantly depressed partial widths, which are absorbed within the ``..." category in Table~\ref{tab:Sigmac1F}. However, it is worth noting that $\Sigma_{c2}(1F,3/2^-)$, $\Sigma_{c2}(1F,5/2^-)$, $\Sigma_{c3}(1F,5/2^-)$, and $\Sigma_{c3}(1F,7/2^-)$ can decay into $\Sigma_{c2}(1D,3/2^+)\pi$, $\Sigma_{c2}(1D,5/2^+)\pi$, $\Sigma_{c3}(1D,5/2^+)\pi$, and $\Sigma_{c3}(1D,7/2^+)\pi$ states, respectively, in an $S$-wave. Consequently, the $\Sigma_c(1D)\pi$ channels play crucial roles in the decay of these states. Besides the $\Sigma_c(1D)\pi$, the $\Sigma_c(1P)\pi$ channels also work in the decays of $\Sigma_c(1F)$. According to the calculations of Refs.~\cite{Chen:2016iyi,Zhou:2023wrf,Yao:2018jmc}, it is potential to observe $\Sigma_c(1P)$ and $\Sigma_c(1D)$ states by the $\Lambda_c\pi$ channel. In this way, the $\Sigma_c(1P)\pi$ and $\Sigma_c(1D)\pi$ then could decay into $\Lambda_c\pi\pi$, which may be an approach to search for the $\Sigma_c(1F)$ states.

In the case of $\Sigma_c(1F)$ states, both the spin and flavor wave functions of the light quarks exhibit symmetry. Hence, in the $\Delta D$ channel, the two light quarks can be considered as a single cluster during the decay process, indicating that the $\Delta D$ channel may have a substantial partial width. For instance, $\Delta D$ accounts for approximately 31.9\% of the branch ratio for $\Sigma_{c4}(1F,7/2^-)$. Since the $\Delta$ could decay into $p\pi$, the observation of $\Delta D$ through $Dp\pi$ could provide a potential avenue for the search of $\Sigma_{c}(1F)$.

Regarding $\Sigma_{c4}(1F,9/2^-)$, significant contributions arise from $\Lambda_c(1P,3/2^-)\pi$ and $\Lambda_c(1D,5/2^-)\pi$ channels. Given that $\Lambda_c(1P,3/2^-)$ and $\Lambda_c(1D,5/2^-)$ are well-established narrow states, searching for the $\Sigma_{c4}(1F,9/2^-)$ state in $\Lambda_c(1P,3/2^-)\pi$ and $\Lambda_c(1D,5/2^-)\pi$ channels would be a viable approach.

\subsection{$\Xi_c^\prime(1F)$ states}

\begin{table*}
\centering
\caption{The partial and total widths of the $\Xi_c^\prime(1F)$ in units of MeV. The conventions are the same as that of Table~\ref{tab:Sigmac1F}.}
\label{tab:Xicp1F}
\renewcommand\arraystretch{1.15}
\begin{tabular*}{178mm}{@{\extracolsep{\fill}}lccccccc}
\toprule[1.00pt]
\toprule[1.00pt]
Decay channels                 &$M_f$ (MeV) &$\Xi^\prime_{c2}(1F,3/2^-)$ &$\Xi^\prime_{c2}(1F,5/2^-)$ &$\Xi^\prime_{c3}(1F,5/2^-)$ &$\Xi^\prime_{c3}(1F,7/2^-)$ &$\Xi^\prime_{c4}(1F,7/2^-)$ &$\Xi^\prime_{c4}(1F,9/2^-)$ \\
\midrule[0.75pt]
$\Xi_c(2S,1/2^+)\pi$           &2970        & 0.7                        & 0.9                        &$\times$                    &$\times$                    &  1.6                       & 1.6                        \\
$\Xi_c(1P,1/2^-)\pi$           &2790        & 0.8                        & 0.0                        & 0.7                        &  0.4                       &  4.7                       & 0.3                        \\
$\Xi_c(1P,3/2^-)\pi$           &2815        & 0.2                        & 0.7                        & 0.8                        &  1.1                       &  1.7                       & 5.7                        \\
$\Xi_c(1D,3/2^+)\pi$           &3055        & 4.3                        & 0.7                        & 7.5                        &  1.1                       &  7.1                       & 0.1                        \\
$\Xi_c(1D,5/2^+)\pi$           &3080        & 0.9                        & 3.2                        & 1.4                        &  6.5                       &  0.6                       & 5.9                        \\
$\Xi^\prime_{c1}(1P,3/2^-)\pi$ &2938        & 0.3                        & 0.3                        & 0.8                        &  1.2                       &  0.2                       & 0.7                        \\
$\Xi^\prime_{c2}(1P,3/2^-)\pi$ &2926        & 0.3                        & 0.3                        & 0.6                        &  0.0                       &  1.4                       & 0.2                        \\
$\Xi^\prime_{c2}(1P,5/2^-)\pi$ &2945        & 0.3                        & 0.4                        & 0.1                        &  0.8                       &  0.4                       & 1.3                        \\
$\Xi^\prime_{c2}(1D,3/2^+)\pi$ &3181        & 5.7                        & 0.3                        & 0.4                        &  0.1                       &  0.2                       & 0.0                        \\
$\Xi^\prime_{c2}(1D,5/2^+)\pi$ &3194        & 0.3                        & 6.1                        & 0.1                        &  0.3                       &  0.0                       & 0.1                        \\
$\Xi^\prime_{c3}(1D,5/2^+)\pi$ &3172        & 0.6                        & 0.1                        & 8.0                        &  0.1                       &  0.7                       & 0.0                        \\
$\Xi^\prime_{c3}(1D,7/2^+)\pi$ &3179        & 0.1                        & 0.6                        & 0.1                        &  8.3                       &  0.1                       & 0.5                        \\
$\Lambda_c(1S,1/2^+)\bar{K}$   &2286        & 0.0                        & 0.0                        &$\times$                    &$\times$                    &  1.2                       & 1.2                        \\
$\Lambda_c(2S,1/2^+)\bar{K}$   &2766        & 1.7                        & 2.1                        &$\times$                    &$\times$                    &  1.7                       & 1.7                        \\
$\Lambda_c(1P,1/2^-)\bar{K}$   &2592        & 2.2                        & 0.0                        & 2.2                        &  1.3                       &  8.5                       & 1.7                        \\
$\Lambda_c(1P,3/2^-)\bar{K}$   &2628        & 0.5                        & 1.6                        & 2.4                        &  3.3                       &  4.3                       &11.1                        \\
$\Lambda_c(1D,3/2^+)\bar{K}$   &2856        & 5.0                        & 1.1                        & 4.5                        &  0.7                       &  1.0                       & 0.0                        \\
$\Lambda_c(1D,5/2^+)\bar{K}$   &2881        & 0.9                        &10.1                        & 0.3                        &  1.9                       &  0.0                       & 0.0                        \\
$\Sigma_c(1S,1/2^+)\bar{K}$    &2455        & 0.1                        & 0.0                        & 0.3                        &  1.4                       &  1.1                       & 0.7                        \\
$\Sigma_c(1S,3/2^+)\bar{K}$    &2520        & 0.2                        & 0.3                        & 2.6                        &  1.9                       &  1.6                       & 2.0                        \\
$\Sigma_{c0}(1P,1/2^-)\bar{K}$ &2788        &$\times$                    &$\times$                    & 1.8                        &  2.0                       &$\times$                    &$\times$                    \\
$\Sigma_{c1}(1P,1/2^-)\bar{K}$ &2766        & 5.4                        & 0.8                        & 4.4                        &  2.6                       &  1.9                       & 0.0                        \\
$\Sigma_{c1}(1P,3/2^-)\bar{K}$ &2798        & 2.4                        &10.4                        & 3.1                        &  4.6                       &  0.3                       & 1.2                        \\
$\Sigma_{c2}(1P,3/2^-)\bar{K}$ &2779        & 4.4                        & 2.3                        &19.4                        &  0.2                       &  3.9                       & 0.7                        \\
$\Sigma_{c2}(1P,5/2^-)\bar{K}$ &2796        & 2.6                        & 5.5                        & 1.8                        & 24.0                       &  0.8                       & 2.9                        \\
$\Lambda_c(1S,1/2^+)\bar{K}^*$ &2286        & 0.2                        & 0.2                        & 1.1                        &  1.2                       &  0.7                       & 0.7                        \\
$\Lambda D$                    &            & 0.0                        & 0.0                        & 0.0                        &  1.9                       &  0.1                       & 4.5                        \\
$\Sigma D$                     &            & 0.0                        & 0.0                        & 0.0                        &  2.9                       &  0.1                       & 6.7                        \\
$\Sigma^* D$                   &            & 5.0                        &30.7                        &16.4                        & 18.0                       & 34.5                       & 5.7                        \\
$\Lambda D^*$                  &            & 0.5                        & 1.0                        & 5.7                        &  4.0                       & 10.6                       & 6.0                        \\
$\Sigma D^*$                   &            & 1.8                        & 3.8                        & 7.4                        &  6.8                       & 10.6                       & 6.0                        \\
$\Sigma^* D^*$                 &            & 3.1                        & 7.4                        & 1.0                        &  3.8                       &$-$                         &$-$                         \\
$\cdots$                       &            & 1.3                        & 1.7                        & 3.8                        &  2.6                       &  2.1                       & 0.4                        \\
\midrule[0.75pt]
Total                          &            &51.8                        &92.6                        &98.7                        &105.0                       &103.7                       &69.6                        \\
\bottomrule[1.00pt]
\bottomrule[1.00pt]
\end{tabular*}
\end{table*}

According to Table~\ref{tab:Xicp1F}, the calculated widths of $\Xi_{c2}^\prime(1F,3/2^-)$ and $\Xi_{c4}^\prime(1F,9/2^-)$ are 51.8 and 69.6 MeV, respectively. The $\Xi_{c2}^\prime(1F,5/2^-)$, $\Xi_{c3}^\prime(1F,5/2^-)$, $\Xi_{c3}^\prime(1F,7/2^-)$, $\Xi_{c3}^\prime(1F,7/2^-)$ may be broad states, which predicted widths are roughly in 90$\sim$100 MeV.

For all six $\Xi_c^\prime(1F)$ states, the partial widths of $\Xi_c^{(\prime)}(1S)\pi$, $\Lambda_c(1S)\bar{K}$, and $\Sigma_c(1S)\bar{K}$ are extremely small. Consequently, it may be challenging to observe $\Xi_c^\prime(1F)$ in these channels. However, our calculations indicate that $\Sigma_c(1P)\bar{K}$ channels may play a significant role for some states. For instance, $\Sigma_{c2}(1P,3/2^-)\bar{K}$ and $\Sigma_{c2}(1P,5/2^-)\bar{K}$ exhibit considerable partial widths for $\Xi_{c3}^\prime (1F,5/2^-)$ and $\Xi_{c3}^\prime (1F,7/2^-)$, respectively. Additionally, according to the calculations in Ref.~\cite{Chen:2016iyi}, the dominant decay mode of both $\Sigma_{c2}(1P,3/2^-)$ and $\Sigma_{c2}(1P,5/2^-)$ is $\Lambda_c\pi$. Therefore, it is possible to observe $\Xi_c^\prime(1F)$ in the $\Lambda_c\bar{K}\pi$ channel. For the $\Xi_{c4}^\prime(1F,9/2^-)$, the $\Lambda_c(1P,3/2^-)\bar{K}$ occupies considerable branch ratio, which provide some clue to search for $\Xi_c^\prime(1F)$.

We also notice that the widths of the $\Sigma^*D$ channel are considerable for some $\Xi_c^\prime(1F)$ states. For example, the branch ratio of $\Sigma^*D$ for $\Xi_{c4}^\prime(1F,7/2^-)$ is approximately 33.3\%. It may be also a approach to search for $\Xi_c^\prime(1F)$ in $\Sigma^*D$ channel.

\subsection{$\Omega_c(1F)$ states}

\begin{table*}
\centering
\caption{The partial and total widths of the $\Omega_c(1F)$ in units of MeV. The conventions are consistent with Table~\ref{tab:Sigmac1F}.}
\label{tab:Omegac1F}
\renewcommand\arraystretch{1.15}
\begin{tabular*}{178mm}{@{\extracolsep{\fill}}lccccccc}
\toprule[1.00pt]
\toprule[1.00pt]
Decay channels                     &$M_f$ (MeV) &$\Omega_{c2}(1F,3/2^-)$ &$\Omega_{c2}(1F,5/2^-)$ &$\Omega_{c3}(1F,5/2^-)$ &$\Omega_{c3}(1F,7/2^-)$ &$\Omega_{c4}(1F,7/2^-)$ &$\Omega_{c4}(1F,9/2^-)$ \\
\midrule[0.75pt]
$\Xi_c(1S,1/2^+)\bar{K}$           &2470        & 0.0                    &  0.0                   &$\times$                &$\times$                &  1.7                   &  1.7                   \\
$\Xi_c(1P,1/2^-)\bar{K}$           &2790        & 0.1                    &  0.7                   &  2.9                   &  1.7                   & 19.2                   &  0.7                   \\
$\Xi_c(1P,3/2^-)\bar{K}$           &2815        & 1.6                    &  1.6                   &  2.9                   &  4.1                   &  5.9                   & 20.7                   \\
$\Xi^\prime_c(1S,3/2^+)\bar{K}$    &2645        & 0.2                    &  0.3                   &  1.1                   &  1.0                   &  0.7                   &  0.9                   \\
$\Xi^\prime_{c0}(1P,1/2^-)\bar{K}$ &2923        &$\times$                &$\times$                &  1.1                   &  1.2                   &$\times$                &$\times$                \\
$\Xi^\prime_{c1}(1P,1/2^-)\bar{K}$ &2899        & 7.0                    &  0.3                   &  2.2                   &  1.3                   &  1.6                   &  0.0                   \\
$\Xi^\prime_{c1}(1P,3/2^-)\bar{K}$ &2938        & 1.9                    & 10.5                   &  1.1                   &  1.7                   &  0.2                   &  0.7                   \\
$\Xi^\prime_{c2}(1P,3/2^-)\bar{K}$ &2926        & 3.5                    &  0.8                   & 18.7                   &  0.0                   &  1.8                   &  0.3                   \\
$\Xi^\prime_{c2}(1P,5/2^-)\bar{K}$ &2945        & 0.8                    &  3.8                   &  1.4                   & 20.5                   &  0.3                   &  1.1                   \\
$\Xi_c(1S,1/2^+)\bar{K}^*$         &2470        & 0.6                    &  0.6                   &  1.2                   &  1.3                   &  0.4                   &  0.4                   \\
$\Xi D$                            &            & 3.4                    &  0.1                   &  2.5                   & 27.9                   &  1.0                   & 64.2                   \\
$\Xi^* D$                          &            &22.5                    & 58.5                   & 61.5                   & 38.9                   &117.3                   & 13.7                   \\
$\Xi D^*$                          &            &30.8                    & 68.5                   & 73.6                   & 77.2                   &110.7                   & 61.3                   \\
$\cdots$                           &            & 0.6                    &  1.1                   &  0.4                   &  0.7                   &  0.6                   &  0.4                   \\
\midrule[0.75pt]
Total                              &            &73.0                    &146.8                   &170.6                   &177.5                   &261.4                   &166.1                   \\
\bottomrule[1.00pt]
\bottomrule[1.00pt]
\end{tabular*}
\end{table*}

According to Table~\ref{tab:Omegac1F}, it can be observed that the $\Xi_c^{(\prime,*)}(1S)\bar{K}$ channels only contribute small fractions to the decay processes. However, in the $\Xi^{(*)}D^{(*)}$  channels, the $c$ quark acts as a spectator while the $ss$ quarks form a cluster. Therefore, it is expected that the $\Xi^{(*)}D^{(*)}$ channels would have considerable partial widths. The numerical results also indicate that the $\Xi D$, $\Xi^*D$, and $\Xi^*D^*$ channels play crucial roles in the decays of $\Omega_c(1F)$ states. Thus, the $\Xi^{(*)} D^{(*)}$ channels are promising channels for the observation of $\Omega_c(1F)$ states.

For the $\Omega_{c2}(1F,3/2^-)$ state, the predicted width is 73.0 MeV. Notably, the $\Xi^* D$ and $\Xi D^*$ channels are identified as two important decay channels with significant branch ratios
\begin{equation}
\begin{split}
{\rm Br}~[\Omega_{c2}(1F,3/2^-)\to \Xi^*D]\approx&30.8\%,\\
{\rm Br}~[\Omega_{c2}(1F,3/2^-)\to \Xi D^*]\approx&42.2\%.
\end{split}
\end{equation}
However, it is worth noting that the predicted widths of $\Omega_{c2}(1F,5/2^-)$, $\Omega_{c3}(1F,5/2^-)$, $\Omega_{c3}(1F,7/2^-)$, $\Omega_{c4}(1F,7/2^-)$, and $\Omega_{c4}(1F,9/2^-)$ states are relatively large, exceeding 100 MeV. This indicates that these states are quite broad compared to the other $\Omega_c(1F)$ states discussed earlier.

\section{Summary}\label{sec:Summary}

In this study, we have conducted an investigation into the spectroscopy behavior of undetected $1F$-wave charmed baryons. By employing a non-relativistic potential model  and the QPC model, we have predicted the masses, widths, and decay channels of these focused states.

Our results indicate that the $1F$-wave charmed baryons exhibit interesting spectroscopic properties. The calculated widths of the different states vary within a certain range, with some states being relatively narrow. These narrow states offer potential opportunities for experimental detection.

We have analyzed the decay modes of the $1F$-wave charmed baryons and identified the dominant and suppressed decay channels. Our findings suggest that certain channels may play significant roles in the decays of these states.
These channels provide promising avenues for the observation of $1F$-wave charmed baryons.

Furthermore, we have discussed the implications of our results for experimental searches. We propose specific channels and decay modes that experimentalists can target in their search for these elusive states, i.e., 
\begin{enumerate}
\item The decay modes involving a $1S$ singly charmed baryon ($\Lambda_c(1S)$, $\Sigma_c(1S)$, and $\Xi_c^{(\prime)}(1S)$) with a pseudoscalar meson ($\pi$ and $\bar{K}$) have small branch ratios for these $1F$ states.

\item The channels that include a light flavor baryon with a heavy flavor meson show potential for observing these $1F$ states.

\item It is recommended to search for $\Sigma_c(1F)$ in the $\Lambda_c\pi\pi$, $\Lambda_c(1P)\pi$, and $\Lambda_c(1D)\pi$ channels, and search for $\Xi_c^\prime(1F)$ in $\Lambda_c\bar{K}\pi$ and $\Lambda_c(1P)\bar{K}$ channels.
\end{enumerate}
By focusing on these suggested channels, it may be possible to detect and study the properties of $1F$-wave charmed baryons, thereby enhancing our understanding of the charmed baryon spectroscopy.

Overall, our study contributes to the ongoing efforts in exploring and characterizing the spectroscopy of charmed baryons, particularly in the $1F$-wave sector. The predicted masses, widths, and decay channels presented here provide valuable insights and guidance for experimental searches and future investigations in this field, especially with the update of high luminosity of LHC.

\section*{ACKNOWLEDGMENTS}

This work is supported by the China National Funds for Distinguished Young Scientists under Grant No. 11825503, the National Key Research and Development Program of China under Contract No. 2020YFA0406400, the 111 Project under Grant No. B20063, the National Natural Science Foundation of China under Grant No. 12247101, the fundamental Research Funds for the Central Universities under Grant No. lzujbky-2022-sp02, and the project for top-notch innovative talents of Gansu province.

\end{document}